# An ECG-on-Chip With 535-nW/Channel Integrated Lossless Data Compressor for Wireless Sensors

Chacko John Deepu, *Member, IEEE*, Xiaoyang Zhang, *Student Member, IEEE*, Wen-Sin Liew, *Member, IEEE*, David Liang Tai Wong, *Student Member, IEEE*, and Yong Lian, *Fellow, IEEE*

*Abstract*—This paper presents a low-power ECG recording system-on-chip (SoC) with on-chip low-complexity lossless ECG compression for data reduction in wireless/ambulatory ECG sensor devices. The chip uses a linear slope predictor for data compression, and incorporates a novel low-complexity dynamic coding-packaging scheme to frame the prediction error into fixed-length 16 bit format. The proposed technique achieves an average compression ratio of 2.25× on MIT/BIH ECG database. Implemented in a standard 0.35 μm process, the compressor uses 0.565 K gates/channel occupying 0.4 mm² for four channels, and consumes 535 nW/channel at 2.4 V for ECG sampled at 512 Hz. Small size and ultra-low-power consumption makes the proposed technique suitable for wearable ECG sensor applications.

## I. INTRODUCTION

HEALTHCARE spending is increasingly becoming the major source of expenditure in many countries, with the U.S. alone spending roughly 18% of its GDP on healthcare. Cardiovascular diseases are one of the leading causes of this spending. Aging and increasing life expectancies are expected to skyrocket these expenses in the near future. The way forward to reign in these costs and improve quality of life is to focus on prevention and early detection of diseases by proactively monitoring the individual's health condition using low-cost wireless wearable sensors.

The main challenge in the development of a low-cost wearable electrocardiogram (ECG) sensor is the design of an ultra-low-power ECG chip, which can acquire, process, and wirelessly transmit the ECG signal to a remote doctor via a personal gateway in real time. A high level of integration, with built-in signal acquisition and data conversion [1], helps reduce the size and cost of such a sensor. The single largest source of power consumption in the sensor is the wireless transceiver. In the scenario of continuous ECG monitoring, a large amount of ECG data is acquired and it has to be either stored locally in a flash memory or transmitted wirelessly to a sensor gateway, resulting in large memory and high energy consumption at the sensor. In some cases, on-chip SRAM blocks are used to buffer the ECG data in order to facilitate burst-mode transmission. However, this results in large chip area [2] and increases the overall cost of the device. Local data compression is an attractive option for such devices. By reducing the amount of data through compression, it helps to minimize the power consumed by the radio for wireless transmission while reducing the size of on-chip SRAM/flash memory and sensor battery. Although lossy compression techniques provide higher compression ratios (CR), we focus on lossless schemes so as to prevent the loss of any signal useful in the diagnostic procedure [3]. Furthermore, lossy compression techniques have not been approved by medical regulatory bodies in many countries and hence cannot be used in commercial devices due to liability concerns. Most of the existing literature on lossless ECG compression predominantly focuses on achieving higher CR. In the context of wireless sensors and ambulatory devices, ultra-low-power operation, low-complexity implementation, and multi-channel support are also important to make sure that the energy and memory savings obtained from the compression are higher than what is consumed by the compressor itself.

This paper describes the development of an ECG acquisition chip with fully integrated lossless compression engine (first presented in [4]) with ultra-low power, which can reduce the system-level power consumption by half without any loss of signal quality. The compression scheme does not require use of costly memory at transmitter or receiver and always generates a convenient fixed-length data output which avoids the need for further packaging. The design has low hardware complexity and achieves low power consumption.

The rest of the paper is organized as follows. In Section II, the system architecture is presented. The analog front-end is detailed in Section III. The compression scheme and performance evaluation are given in Section IV. The architecture of the proposed scheme and its implementation are discussed in Section V. Measurement results are shown in Section VI. Conclusions are given in Section VII.

Manuscript received January 28, 2014; revised May 23, 2014; accepted July 02, 2014. This paper was approved by Guest Editor Zhihua Wang. This work was supported in part by the National Research Foundation Competitive Research Programme under Grant NRF-CRP8-2011-01 and by the Faculty Strategic Research Program under Grant R-263-000-A02-731.

C. J. Deepu, X. Zhang, and D. L. T. Wong are with the BioElectronics Laboratory, Department of Electrical and Computer Engineering, National University of Singapore, Singapore 117576.

Y. Lian is with the School of Microelectronics, Shanghai Jiao Tong University, Shanghai 200240, China, and also with the Department of Electrical and Computer Engineering, National University of Singapore, Singapore 117576 (e-mail: eleliany@gmail.com).

W.-S. Liew was with the BioElectronics Laboratory, Department of Electrical and Computer Engineering, National University of Singapore, Singapore, and is now with Avago Technologies, Singapore.

Color versions of one or more of the figures in this paper are available online at http://ieeexplore.ieee.org.

Digital Object Identifier 10.1109/JSSC.2014.2349994

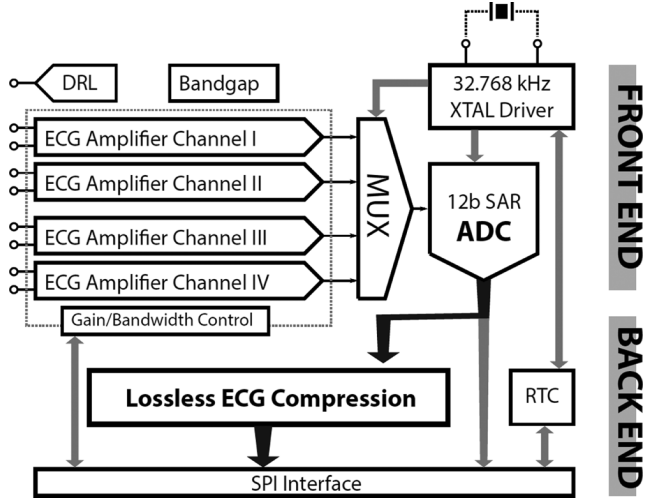

Fig. 1. System architecture of the ECG SoC.

## II. SYSTEM ARCHITECTURE OF ECG SoC CHIP

The system block diagram of the proposed ECG SoC is shown in Fig. 1. The front-end consists of four ECG recording channels, a multiplexer (MUX), and a 12 bit successive approximation (SAR) ADC. The digital back-end includes a lossless compression block, a real-time clock (RTC) module, and a serial peripheral interface (SPI). To improve the ECG signal quality and suppress the 50/60 Hz power-line interference, a driven-right-leg (DRL) circuit is included. The output of the DRL is connected to the right-leg (RR) electrode to stabilize the subject potential and improve the common-mode suppression. A low-power 32.768 kHz crystal oscillator driver and a CMOS bandgap reference are also integrated on-chip in order to minimize the number of off-chip auxiliary circuits. The ADC sampling rate is configurable for either 256 or 512 Hz for a balance between signal quality and the amount of data. The whole chip is designed to work under a single 2.4 to 3.0 V power supply.

## III. ANALOG FRONT-END

The analog front-end (AFE) is often a bottleneck of the system in terms of noise and linearity performance. The input-referred noise needs to be low enough for accurate biomedical data acquisition. The signal distortion should be less than 1% even at the 3 V full-scale output. Moreover, as the sampling rate of the ADC is at 256/512 Hz, a higher order low-pass filter with less than 100/200 Hz cut-off frequency is required for minimizing the aliasing errors. The limited power budget discourages extra active anti-aliasing filters. In our design, the signal bandwidth is reduced by designing a low-bandwidth operational transconductance amplifier (OTA) for both the instrumental amplifier (IA) and the programmable gain amplifier (PGA). This section highlights the circuit design considerations and trade-offs for the analog front-end.

### A. ECG Channel With Pseudo Resistors

Fig. 2 shows the architecture for a single-channel ECG front-end. The AFE includes a low-noise IA, a PGA, and a rail-to-rail output buffer (BUF). The IA amplifies the ECG signal with a fixed gain of 125. Tunable pass-band gain is achieved by tuning the PGA gain through G$\langle 1{:}0 \rangle$. For anti-aliasing purposes, the low-pass cut-off frequency can be adjusted within 35–175 Hz, by changing the PGA's frequency response. The following unity-gain buffer improves the settling time for the MUX output signal, reducing the residual errors [1].

The IA adopts the capacitively coupled technique to block the input DC offset. While the amplitude of the typical ECG is around millivolts, the DC offset between the differential Ag/Cl wet electrodes could be up to 200 mV, or even higher when using dry electrodes. To avoid saturating the amplifiers and to increase the input dynamic range, the input offset shall be cancelled properly.

A simple and power-efficient way to block the DC offset is by using a high-pass filter. Since the low-frequency component of the ECG traces around 0.5 Hz still contain important information for ST segment analysis, the high-pass corner needs to be set at 0.05 Hz or lower. Large capacitors and resistors are hence required to achieve such a low cut-off frequency. In our design, pseudo resistors [5] replace the passive resistors to save the chip area. The pseudo resistors are normally constructed by two or more diode-connected PMOS in parallel. The bulk of each PMOS is connected to the source or drain, so that the performance of the pseudo resistor is less dependent on the absolute voltage applied onto it.

Two types of pseudo resistors are used in the IA and the PGA stage respectively. The simulated resistance versus input voltage across the pseudo resistors is plotted in Fig. 3. As $C_1$ in the IA stage is around 0.5 pF, the equivalent resistance of the pseudo resistor in this stage should be at least $6.4 \times 10^{12}\ \Omega$. Because the input amplitude is small for the IA stage, it is unnecessary for the pseudo resistor in the IA stage to support large input. So the Type A design with two PMOS transistors is adequate. For the PGA stage, however, the input amplitude could be as high as 1.5 V. Since the resistance of the Type A design is around 1 M$\Omega$ at $\pm 1.5$ V, the current flowing through the pseudo resistor would be about 1 $\mu$A when output amplitude is large, causing loading errors at the output. It is therefore necessary to use a different pseudo resistor structure with higher resistance for a wide input range. By cascading two Type A designs in series, the resulting resistance of the Type B design is $10^5$ higher than Type A at 1.5 V. As shown in the testing result later in Section VI, the cascaded pseudo resistor helps to achieve less than 0.4% total harmonic distortion (THD) at 3 V output.

### B. Operational Transconductance Amplifier (OTA)

Fig. 4 shows the schematic of the low-noise OTA used in the IA for ECG capturing. The fully-differential configuration increases common-mode suppression, making the ECG signal less affected by common-mode artifacts and power-line interference. Two-stage architecture is used to improve output swing and open-loop gain. At the first stage, an extra $g_m$ branch with $M_1$–$M_4$ as input transistors is added. The branch consists of an inverter-based differential pair, which improves the $g_m/I$ efficiency. By tuning the current between $M_1$-$M_4$ and another

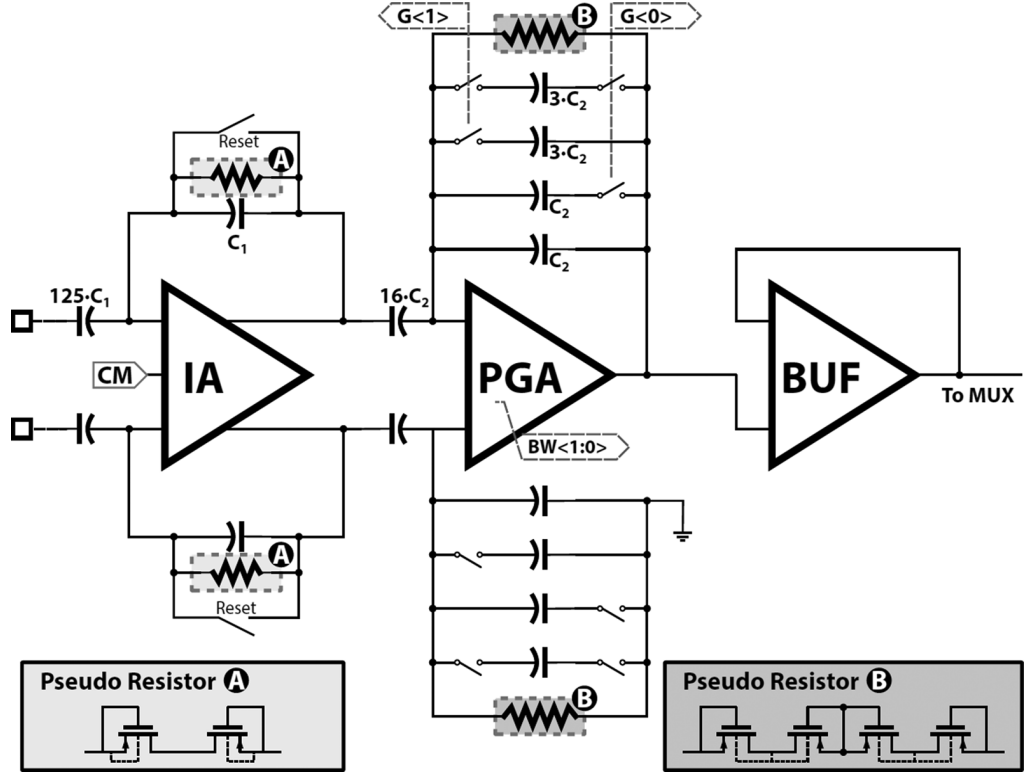

Fig. 2. Architecture of a single ECG amplifier channel.

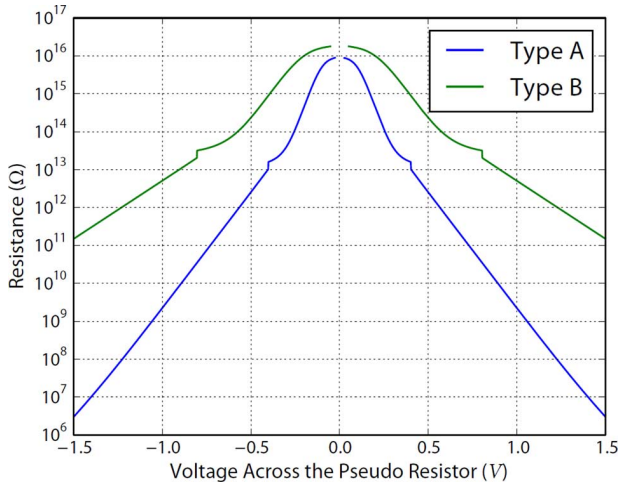

Fig. 3. Simulated resistances for different type of pseudo resistors.

input pair, $M_5$ and $M_6$, the OTA transconductance $g_{mi}$ can be changed, which is given by

$$g_{mi} = g_{m1,2} + g_{m3,4} + g_{m5,6}. \quad (1)$$

To improve the noise-to-power efficiency, all of the input transistors $M_1$–$M_6$ are biased in the subthreshold regime. The thermal noise current of a MOS transistor operated in the weak inversion [6] can be modeled as

$$\overline{i_{n,\text{thermal}}^2} = 2kTn \cdot g_m \quad (2)$$

where $n$ is the subthreshold slope factor, which is around 1.3 as simulated in the target technology. The noise contributions of the cascaded transistors and the tail current sources are negligible. Also, the first stage dominates the noise for a typical two-stage OTA. Based on the simplification mentioned, the input-referred thermal noise of this OTA is approximately

$$\overline{v_{ni,\text{thermal}}^2} \approx \frac{4kTn}{g_{mi}} \left(1 + \frac{g_{m7,8}}{g_{mi}} + \frac{g_{m9,10}}{g_{mi}}\right) \Delta f. \quad (3)$$

The noise efficiency factor (NEF) [7] is used to benchmark the noise-to-power trade-off, which is defined by

$$\text{NEF} = v_{ni,\text{rms}} \cdot \sqrt{\frac{2I_{\text{tot}}}{\pi \cdot U_T \cdot 4kT \cdot \text{BW}}} \quad (4)$$

where $I_{\text{tot}}$ is the total current and BW is the amplifier bandwidth. Suppose that the drain current $I_d$ of $M_1$–$M_4$ is $\alpha \cdot I_{\text{tot}}$, and the drain current of $M_5$ and $M_6$ is $\beta \cdot I_{\text{tot}}$. Also under the EKV model [8], the $g_m$ of a subthreshold MOS transistor is approximately

$$g_m = \frac{I_d}{nU_T} \quad (5)$$

where $U_T = kT/q \approx 26$ mV. If we further ignore the flicker noise and the noise contributions from $M_7$–$M_{10}$ and the common-mode feedback (CMFB) circuit, the optimal NEF for this OTA architecture is given by

$$\text{NEF}_{\text{opt}} = \sqrt{\frac{n^2}{2\alpha + \beta}}. \quad (6)$$

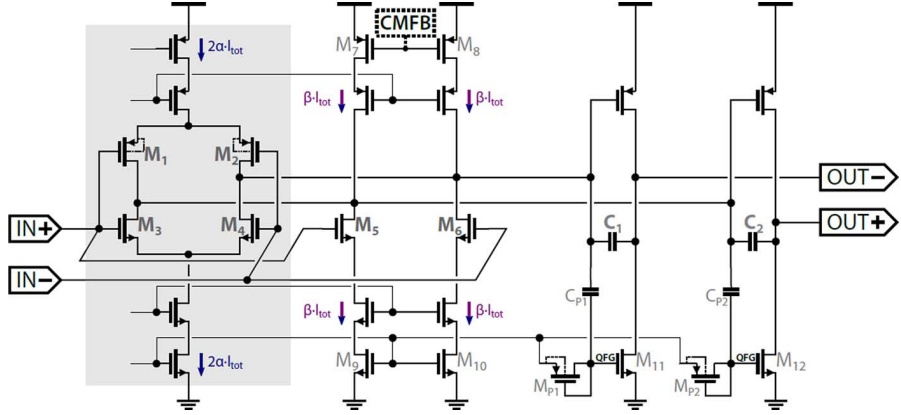

Fig. 4. Instrumental amplifier OTA used in ECG channels.

Nonetheless, it should be noted that the current of the first stage cannot exceed the total current $I_{\text{tot}}$. To minimize the NEF, given that $2\alpha + 2\beta < 1$ is required, $\alpha$ should be maximized to take advantage of the current reuse between $M_1$–$M_4$. If $\alpha = 0.5$ that all the current flows to the inverter-based amplifier branch, the minimum NEF of 1.3 can be achieved. Unfortunately, pursuing the highest NEF would impair other design targets like the bandwidth limits and the settling time, which is discussed in the following.

First, unlike other high-speed designs, it is advantageous to limit the bandwidth of the IA for ECG capturing. The intrinsic low-pass characteristic of the OTA helps to suppress the high-frequency noise and artifacts without any extra active filters. This issue becomes even more critical as the sampling rate is 256/512 Hz or lower. Since the bandwidth of the IA is

$$\text{BW} = \frac{g_{mi}}{C_{1,2} \cdot G_{\text{IA}}} \quad (7)$$

both the IA pass-band gain, $G_{\text{IA}}$, and the Miller-compensation capacitance values, $C_1$ and $C_2$, have to be increased to limit the bandwidth. As the gain is also determined by the $C_1$ capacitor ratio given in Fig. 2, either approach demands significant capacitor area on chip. Alternatively, the $g_{mi}$ can be reduced, with the side effect of increasing the noise floor.

Second, large output slew rate is required to mitigate the output distortion. Because the IA gain $G_{\text{IA}}$ is designed to be 125, harmonic distortions could be introduced even at the IA stage. The most straightforward way to improve the linearity is to increase the static current at the output stage. Simulation also shows large output stage current improves the baseline recovery time after resetting the IA. But excessive current at the second stage would inevitably affect the power utilization and the transconductance $g_{mi}$. An output boosting technique called quasi-floating gating [9] is also used to push the second stage of the OTA into class-AB operation. The gates of $M_{11}$ and $M_{12}$ are partially controlled by the first stage's outputs through the small capacitors $C_{P1}$ and $C_{P2}$, so that the transconductance of the second stage $g_{m11}$ and $g_{m12}$ is enhanced.

Last but not least, the common-mode feedback circuit should be carefully designed to avoid stability issues. As the drain currents of $M_5$ and $M_6$ are controlled by the common-mode feedback circuit, setting $\beta$ too small would cause the CMFB to fail to adjust the common-mode current. On the other hand, using large $\beta$ is likely to introduce CMFB stability issues. Moreover, the CMFB circuit itself requires minimal current dissipation to ensure enough common-mode settling time and the loop stability.

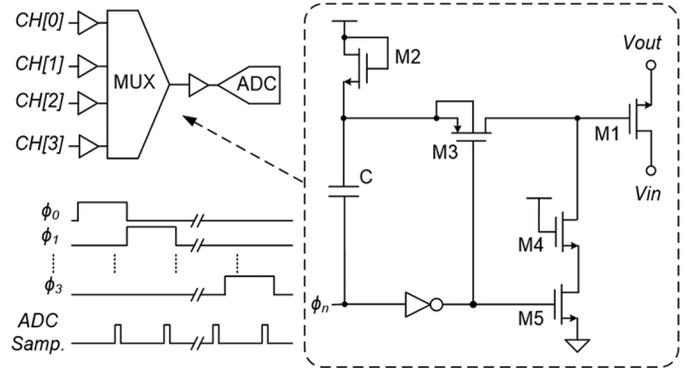

Fig. 5. Schematic and timing of analog MUX and ADC.

With all the trade-offs mentioned above, we allocate half of the total current to the second stage to improve the output linearity and the baseline recovery time after reset. The current ratio $\alpha$ and $\beta$ are both set at 0.1, considering the bandwidth upper limits. The optimal NEF now becomes 2.4. It is worth noting that this simplified calculation does not include the flicker noise, which is optimized by using large width and length for all input transistors, $M_1$ to $M_6$.

### C. Multiplexer (MUX) and Analog-to-Digital Converter (ADC)

All channels are multiplexed to the ADC for AD conversion using an analog MUX. The ADC is based on the dual-capacitive-array architecture proposed in [10]. The MUX is implemented using bootstrapped switches as shown in Fig. 5. The bootstrapped technique allows reduced-size NMOS to be used as the switch. Consequently, the associated parasitic capacitance and on-resistance of the switches are much smaller as compared to the conventional transmission gate. This not only ensures that the system bandwidth is not limited by the multiplexer but also minimizes the signal interference between channels. Since the breakdown voltage is 5 V in the target technology, device M2 is chosen as a diode-connected NMOS. With NMOS threshold voltage of about 1 V, this limits the boosted voltage to less than 5 V under a 3 V supply. As shown in the timing diagram, the MUX multiplexes four inputs to the ADC in sequential order. Its switching is misaligned with ADC sampling and sufficient settling time is assigned before ADC sampling to minimize signal distortion. Buffers are inserted before and after the MUX in order to address the driving issue.

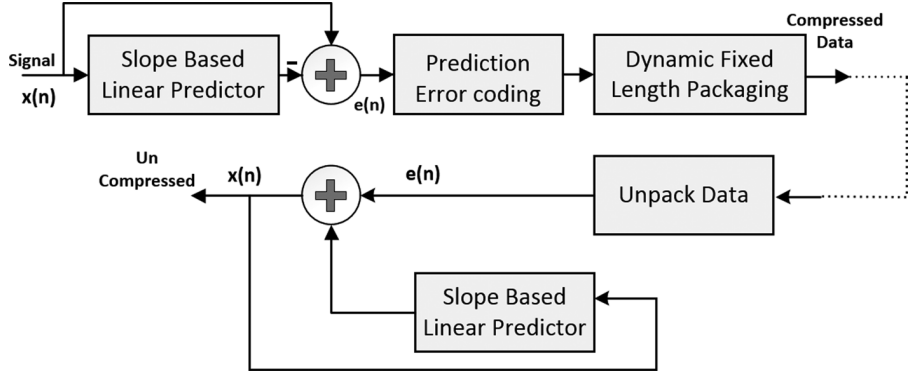

Fig. 6. Lossless compression–decompression scheme.

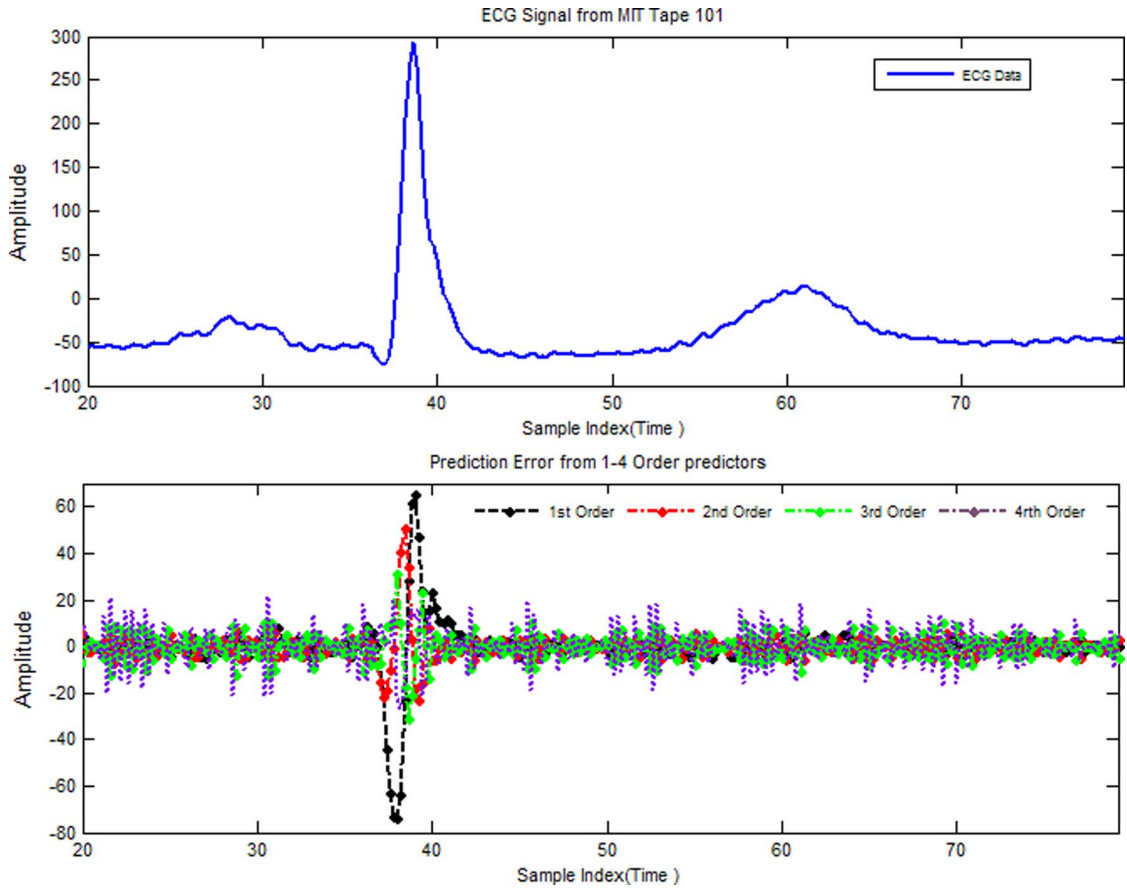

Fig. 7. Error from various predictors.

## IV. LOSSLESS DATA COMPRESSION SCHEME

The block diagram of the proposed compression-decompression scheme is illustrated in Fig. 6. A short-term linear predictor is used to model the ECG and de-correlate the input signal. Here, the current sample value, $x(n)$, is estimated from past samples:

$$\hat{x}(n) = \sum_{k=1}^{L} a_k x(n-k) \qquad (8)$$

where $\hat{x}(n)$ is the estimate of the sample $x(n)$, $a_k$ is the predictor coefficient, and $L$ is the order of the predictor. Knowing that an ECG signal has minimal variation between samples and most signal energy is in the low frequencies, differentiators of orders ranging from 1 to 4 with integer coefficients are tested using MIT/BIH data to select an optimal predictor with minimal complexity. The coefficients of the 1st to 4th order predictors are $[1]$, $[2, -1]$, $[3, -3, 1]$, $[4, -6, 4, -1]$, respectively. The predicted value is then subtracted from the actual value to obtain the prediction error, i.e., $e(n) = x(n) - \hat{x}(n)$, of the current sample. For optimal compression, the prediction error should be as low as possible. Fig. 7 shows the prediction errors from MIT/BIH tape 101 for various predictors. It is observed from Fig. 7 that, due to the variation in signal statistics, predictors perform differently for various segments of the ECG signal. For segments with large amplitude variation, higher order predictors perform better, and for slow varying segments, lower order predictors perform better. Since large amplitude variations are mostly limited to the QRS segment, for which the duration is less than ~12% of one ECG cycle, lower order predictors are expected to have better average prediction performance. To verify this and find the optimal predictor with the highest average performance, mean absolute prediction error (MAPE) and root-mean-square prediction error (RMSPE) for four predictors are computed for all the records in the MIT/BIH Arrhythmia database using (9) and (10).

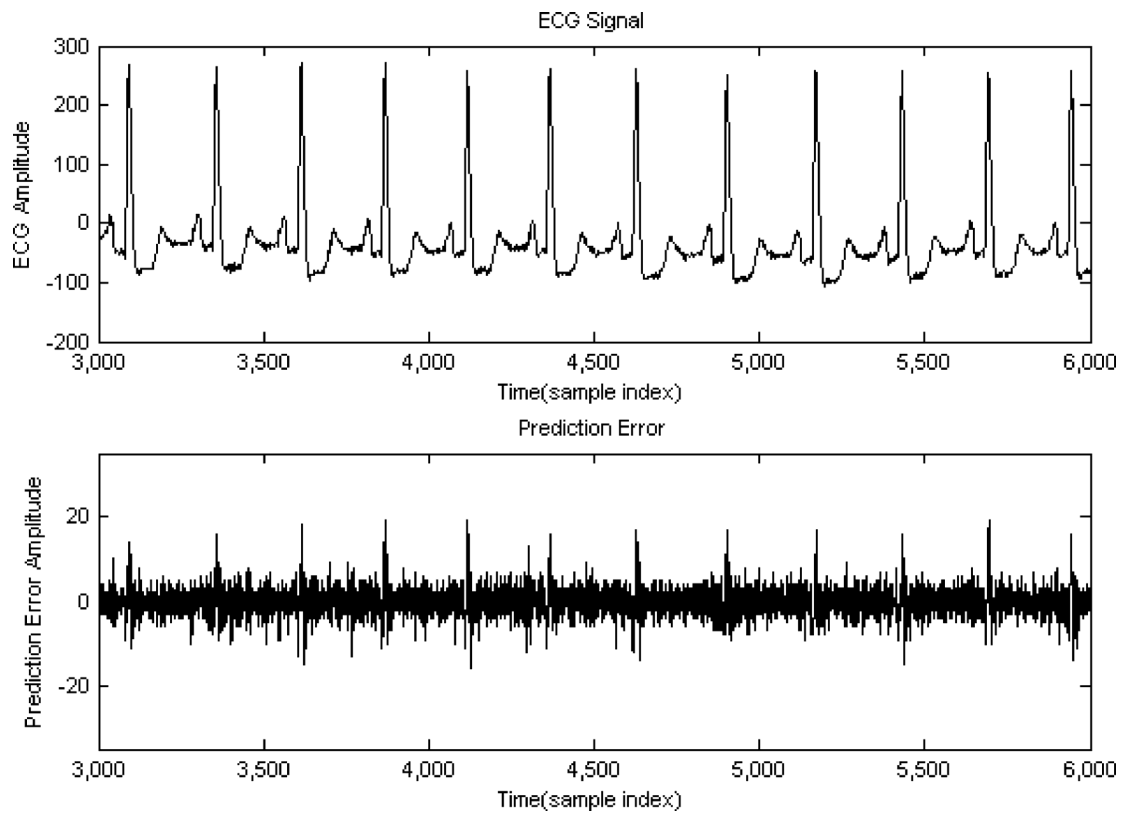

Fig. 8. Dynamic range of the signal after de-correlation (MIT Tape 105).

$$\text{MAPE} = \frac{1}{N} \sum_{n=1}^{N} |x(n) - \hat{x}(n)| \quad (9)$$

$$\text{RMSPE} = \sqrt{\frac{1}{N} \sum_{n=1}^{N} |x(n) - \hat{x}(n)|^2} \quad (10)$$

TABLE I
MEAN AND ROOT MEAN SQUARE PREDICTION ERROR
FOR MIT/BIH TAPES 104 AND 203

| Predictor | MIT/BIH Tape 104 | | MIT/BIH Tape 203 | |
|---|---|---|---|---|
| | MPE | MSPE | MPE | MSPE |
| 1st-order predictor | 4.46 | 11.001 | 6.37 | 11.03 |
| 2nd-order predictor | 3.95 | 9.19 | 4.53 | 6.71 |
| 3rd-order predictor | 5.44 | 10.52 | 5.91 | 8.22 |
| 4th-order predictor | 8.75 | 14.39 | 9.41 | 12.65 |

where $N$ is the number of samples in the ECG record. The second-order predictor yields the lowest overall MAPE and RMSPE and therefore it is chosen for de-correlation of the ECG signal. The MAPE and RMSPE of two sample ECG records, e.g., tapes 104 and 203, are given in Table I to show the performance differences. After de-correlation, the dynamic range of $e(n)$ is much smaller than the ECG signal, as shown in Fig. 8. Note that, for achieving lossless compression, it needs a maximum of $M + 2$ bits to fully represent $e(n)$, where $M$ is the bit width of $x(n)$. With the proposed scheme, only prediction error needs to be transmitted instead of the original ECG samples. At the receiver, the exact reverse process has to be carried out to reconstruct the original data as in Fig. 6.

To reduce the bit width of $e(n)$, a coding scheme can be used without incurring any data loss. *Variable-length coding* schemes like Huffman and Arithmetic coding [11] are commonly used, which produces *prefix-free* codes [12] that can be packed closely. Though these approaches produce relatively optimal bit representations, the complexity of the encoder and decoder is quite high [11], [13]. For example, the Huffman coding method associates the most frequently occurring symbols with short codewords and the less frequently occurring symbols with long codewords. This symbol–codeword association table has to be pre-constructed using a statistical dataset. The implementation of this table, for a fully lossless compression, would require a large on-chip memory (i.e., $2^{13}$ locations for a 13 bit $e(n)$) [13], which will eventually compromise the savings of SRAM area achieved by the use of data compression. A suboptimal approach [12], selective Huffman coding, encodes only $m$ frequently used symbols with Huffman codes and retains the remaining data unencoded at the expense of a decrease in compression ratio. The hardware complexity of [12] is lower compared to the statistical approach. However, it still needs an $m$-symbol lookup table at the encoder, as well as the decoder. In addition, these coding schemes produce variable-length codes at the output and thus require further packaging into fixed-length packets before it can be stored in fixed-word-length SRAM/flash memory or interfaced through a standard I/O like SPI. This repackaging usually needs complex hardware like that proposed in [14].

We propose a simple coding-packaging scheme, which combines encoding and data packaging in one single step. It has very low hardware complexity and achieves small area and low power while producing a fixed-length 16 bit output. The flowchart of the scheme is presented in Fig. 9, where the error signal is represented in 2's complement format ($e\_2c(n)$). As shown in Fig. 8, most error samples center on zero and hence can be represented by a few bits. Therefore, we only select the necessary LSBs and remove any MSBs that do not carry any information. However, these data cannot be packed closely because they do not have the *prefix-free* nature of Huffman codes. Consequently, a data framing structure, shown in Table II, with a unique header for different frame types, is formed to pack the error samples of varying widths to a fixed-length 16 bit output.

The dynamic data packaging scheme, as shown in Fig. 9, uses a priority encoding scheme to frame fixed-length data from samples of multiple bit widths. As the error data is received, the algorithm checks whether the maximum amplitude of a group of the last several signal samples exceeds the value that a particular frame type can accommodate from Table II. If not, the algorithm proceeds with the next best framing option. The order of priority of frame generation is D, C, A, B, E. For Type E frames, the original sample itself is sent instead of prediction error.

### A. Performance Evaluation

The bit compression ratio (BCR) of the scheme is estimated as in (11). It indicates the number of uncompressed bits corresponding to each compressed bit.

$$\text{BCR} = \frac{\text{No. of uncompressed samples} * \left(\frac{\text{bits}}{\text{sample}}\right)}{\text{No. of compressed samples} * \left(\frac{\text{bits}}{\text{sample}}\right)}. \quad (11)$$

The proposed algorithm is evaluated using ECG records from 48 patients in the MIT/BIH Arrhythmia database, sampled at

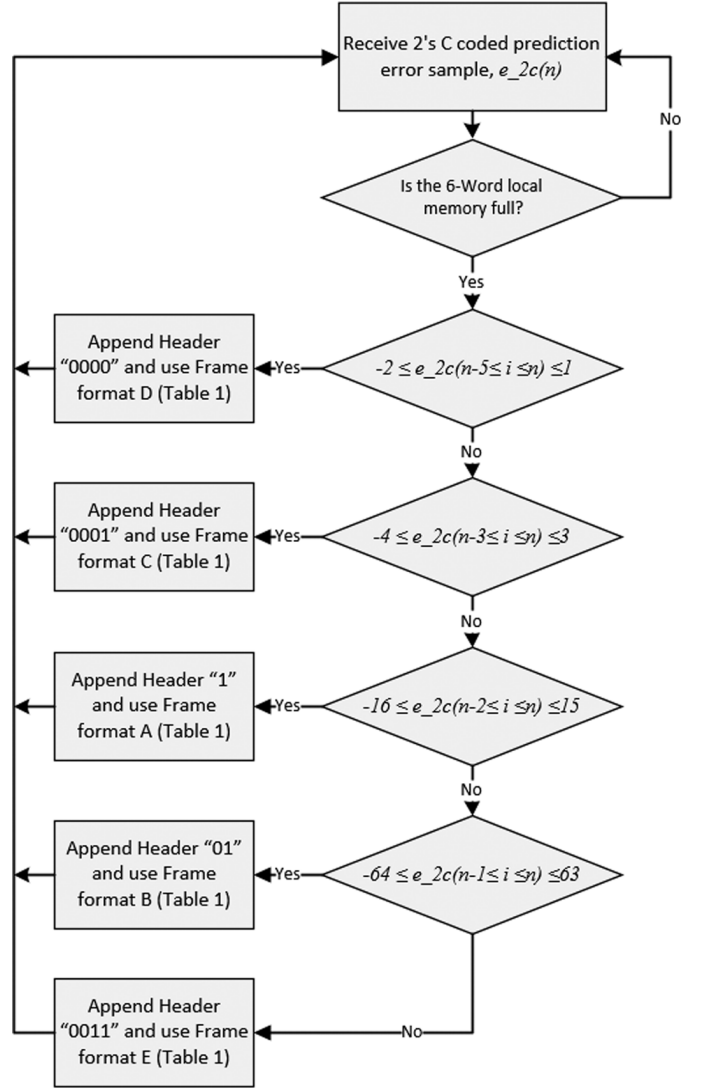

Fig. 9. Coding-packaging scheme flowchart.

TABLE II
DATA PACKAGING SCHEME FOR PREDICTION ERROR SYMBOLS

| | | | | | | | | | | |
|---|---|---|---|---|---|---|---|---|---|---|
| A | 1 | | | 5/4 bits | | 5/4 bits | | 5/4 bits | | |
| B | 0 | 1 | | 7/6 bits | | | | 7/6 bits | | |
| C | 0 | 0 | 1 | 3 bits | | 3 bits | | 3 bits | | 3 bits |
| D | 0 | 0 | 0 | 2 bits | 2 bit | 2 bits | | 2 bits | | 2 bits |
| E | 0 | 0 | 1 1 | 12 bits | | | | | | |

| Frame Type | Header | Data |

360 Hz [15]. An average compression ratio of 2.25 is achieved against all the records. The compression performance of the proposed algorithm with MIT/BIH Arrhythmia database is given in Table III and Fig. 10. The proposed coding scheme achieves 4% better performance than that of Selective Huffman coding, while generating fixed-length coded output. Its compression ratio is around 15% lower than ideal Huffman coding, but at significantly less hardware cost, as will be discussed in the next section. To ensure that the proposed scheme can be applied to any ECG dataset, we also tested the scheme with the MIT/BIH compression test database, which consists of 168 patient records

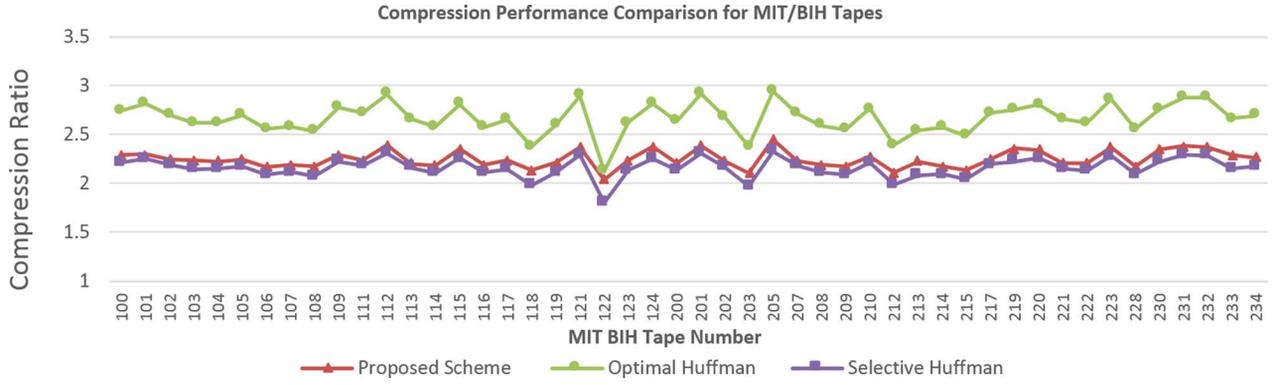

Fig. 10. Comparison of prediction error for 48 MIT/BIH records.

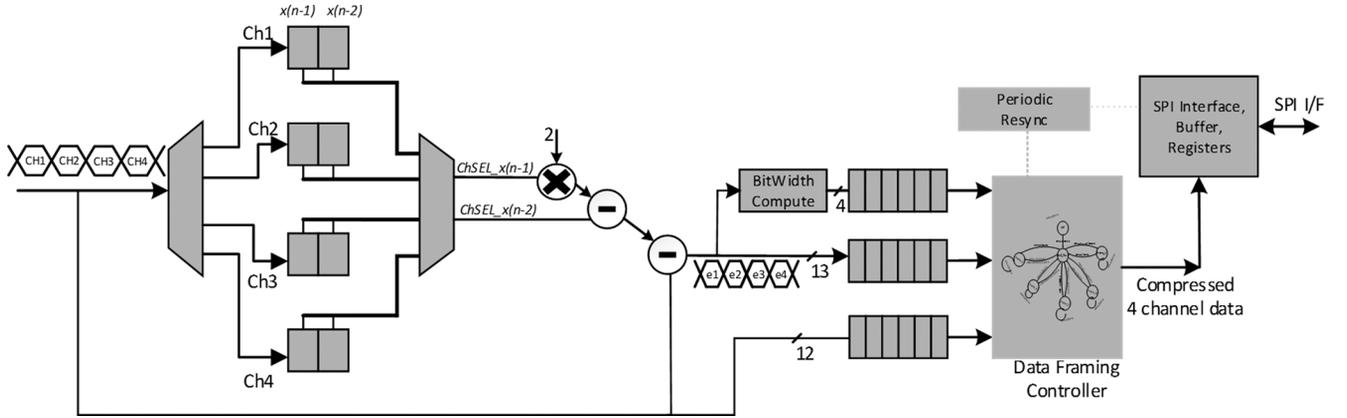

Fig. 11. Architecture for a 4-channel lossless ECG compressor for ECG on chip.

TABLE III
COMPRESSION PERFORMANCE OF THE PROPOSED TECHNIQUE
USING MIT/BIH DATABASE

| Tape | Slope Predictor + Ideal Huffman | Slope Predictor + Selective Huffman | Slope Predictor + Proposed Scheme |
|---|---|---|---|
| Avg. BCR | 2.66 | 2.15 | 2.25 |
| Max. BCR | 2.94 | 2.32 | 2.44 |

sampled at 250 Hz [16]. An average compression ratio of 2.198 is achieved for all the patients.

## V. ARCHITECTURE FOR THE PROPOSED COMPRESSOR

An overview of the hardware architecture for a 4-channel compressor is shown in Fig. 11. The design takes 12 bit, multiplexed 4-channel data from ADC output. The incoming data is serially multiplexed in the format of $\langle ch1, ch2, ch3, ch4 \rangle$. Since the compressor works on each channel data independently, de-multiplexing has to be performed before further processing the data. The data from each channel is identified by a 2 bit channel select header appended to the ECG sample by the ADC. Each of the data streams is fed into a separate slope predictor for computing the prediction error. The individual predictors are gated with the channel-select signal to reduce the effective data switching. Since the incoming data is serially multiplexed while the slope predictor requires two consecutive samples from the same channel to produce the linear prediction, we use buffers to form two consecutive samples from serially multiplexed input stream. This way, the same arithmetic hardware for computing the linear prediction can be shared among all channels.

When initializing the compression operation, the first two samples of all channels have to be explicitly transmitted or predetermined. This is to aid the initialization of the decompression process at the receiver. In our design, we predetermined the first two samples of all channels as zeroes, i.e., predictor data registers were set to zeroes during initialization. As a result, initial samples need not be explicitly transmitted, which in turn simplifies the hardware implementation.

The next step is to compute the minimum bit-width required for representing each prediction error sample, $e(n)$, in 2's complement format. The bit width required for each error sample is computed and loaded into a 6-word register. Accordingly, the error amplitude comparison operations required for data framing can be reduced into comparing the bit widths of $e(n)$. Since the data framing format shown in Table II only accepts five types of data packets (i.e., 2, 3, 5, 7, and >8 bits), the bit widths of $e(n)$ with only these granularities are computed by the *BW compute n bit* block, as shown in Fig. 12(a), which checks whether the incoming sample can be represented by using $n$ bits. A priority encoder is used to find the absolute minimum bit width required for representing the sample. The encoder assigns the highest priority to the block output, which assigns the lowest required bit width. The encoding table is shown in Fig. 12(c).

The schematic of the bit-width computation logic is illustrated in Fig. 12(b). The circuit checks if the leading $(12 - n)$

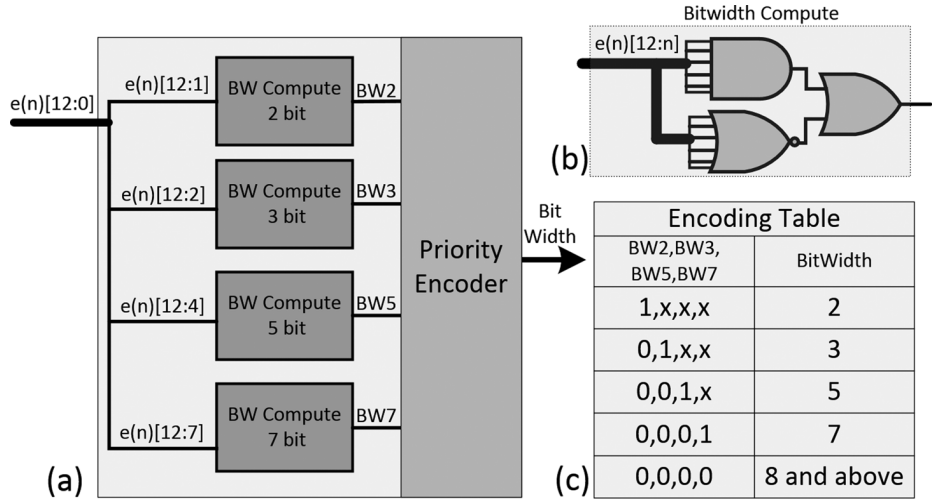

Fig. 12. (a) Bit-width computation block. (b) Bit-width computation logic. (c) Encoding table.

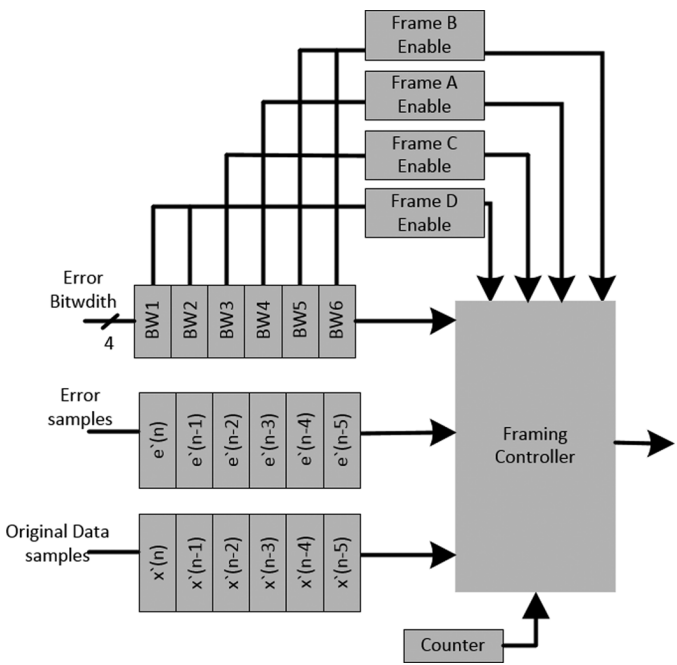

Fig. 13. Data framing block logic.

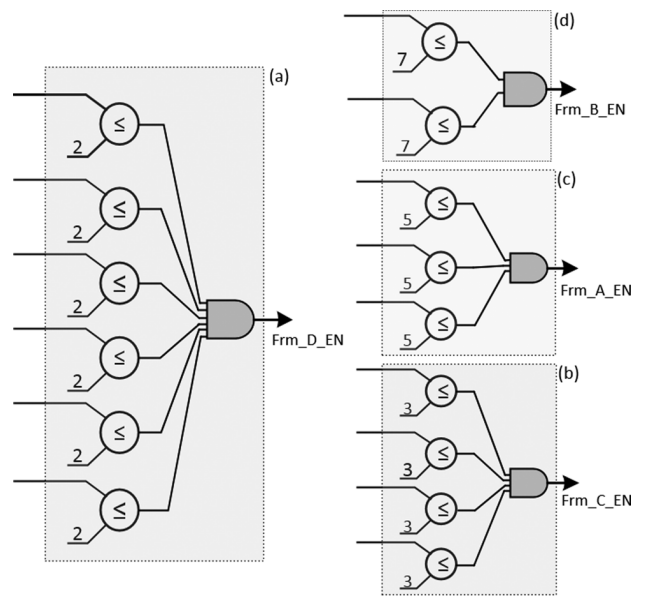

Fig. 14. Frame-enabling logic for (a) Type D, (b) Type C, (c) Type A, and (d) Type B.

MSBs of the error sample, represented in 2's complement, is either '0' or '1'. If so, this indicates that the sample can be represented in $n + 1$ bits.

The error samples are then packaged into fixed 16 bit frames, based on the minimum required bit widths of the error samples. The overview of the framing block hardware is given in Fig. 13. The original data samples, error samples, and corresponding minimum bit-widths required are loaded into a 6-word register. Once the register is full, the framing controller packs the data from the register into a single 16 bit frame based on the input from frame-enabling logic. The frame-enabling logic is used to evaluate if the current samples in the register could be packed into one of the five frame types shown in Table II. This is implemented using simple comparators which compare the bit-width values in the register with those required for each frame type. The implementation of frame-enabling logic is shown in Fig. 14. The framing controller assigns the highest priority to frame Type D and lowest to frame Type E (resynchronization frame).

### A. Automatic Resynchronization

Similar to other schemes in the literature [11], [17], the reconstructed signal will be in error and cannot be recovered if there is a packet loss in the wireless channel. To mitigate this issue, we devised a simple technique wherein we periodically send resynchronization frames (i.e., frame Type E), so that even if a packet loss occurs, the decompression can be restored by the time the next resynchronization frame appears. These frames contain the original samples, not the prediction error, and hence the receiver is able to resynchronize the register values of the decompression block, even in the case of packet loss. In the case of a packet loss, the maximum amount of data lost will be limited to the time between the packet loss and the appearance of the next resynchronization frame, as illustrated in Fig. 15(a). The resynchronization frames are transmitted every 4 seconds in this design. The selection of a 4 s interval for resynchronization is based on observations made in past, i.e., the corrupted data is limited to less than 1% if there is a packet loss every 10 minutes [18]. Certainly, this interval can be adjusted based on transceiver performance. In case of a much higher rate of loss, error correction schemes provided by the transceiver or higher level storage and retransmission mechanisms should be enabled. This is true even for transmission of noncompressed biomedical data, as data loss is generally unacceptable.

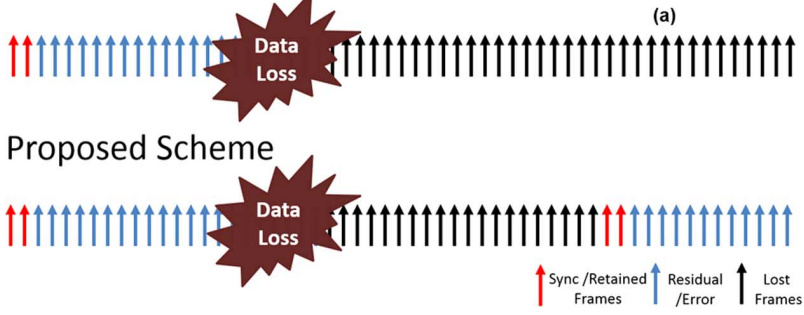
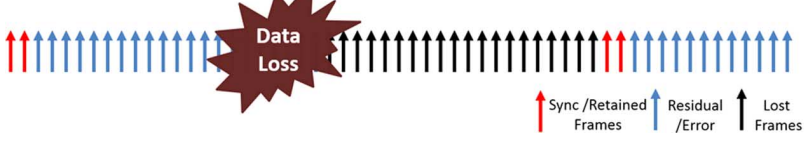
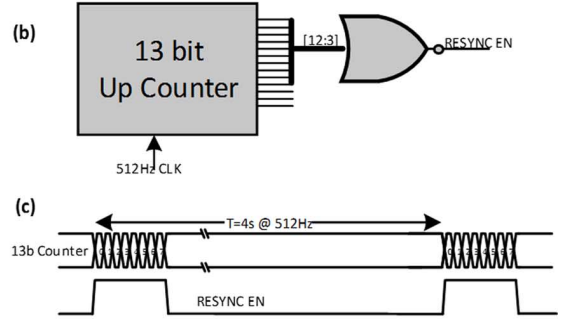

Fig. 15. (a) Illustration of resynchronization. (b) Control signal block. (c) Resynchronization control signal.

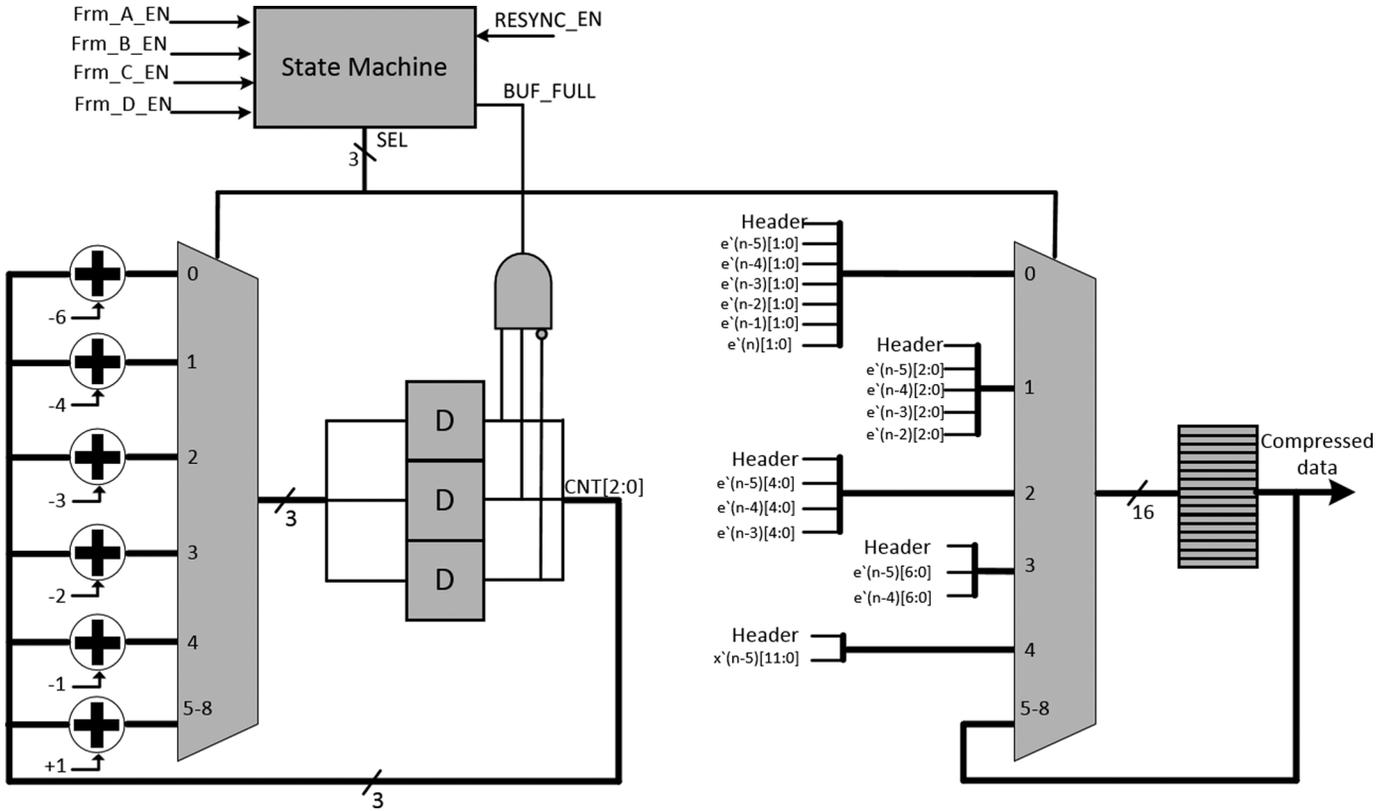

Fig. 16. Schematic of the framing controller.

The control signal for enabling the resynchronization, *RESYNC_EN*, is generated for the framing controller as shown in Fig. 15(b) and (c). In typical laboratory conditions, using the proposed chip, such errors never occurred.

### B. Framing Controller

The framing controller block creates 16 bit data frames based on the input from the prediction error. To keep track of the number of valid samples in the registers after each framing operation, a 3 bit counter, as shown in Fig. 16, is used to indicate that registers are fully loaded and ready for a framing operation if its value reaches 6. The state machine for the controller is shown in Fig. 17.

At reset, the state machine is in state *INIT* and the counter is incremented by '1' for every data loaded into the register. The state machine output asserts the multiplexer select signal *SEL* (Fig. 16) to '5' which enables the counter to count up by '1' through the corresponding multiplexer input. Once the counter reaches '6', the state machine goes to state *BUF_FULL* and start to identify optimal framing options starting from frame Type D, which can pack up to 6 samples at a time. The state machine uses control signal *CTRL* (Fig. 17) which is a concatenated 5 bit signal from the framing enabling logic and *RESYNC_EN* signals. If $CTRL = \langle 1, x, x, x, 0 \rangle$, where $x$ refers to "don't care"; it indicates that Type D frames can be formed from the current register data. The state machine goes to state *Frame_D*, asserts the *SEL* signal to '0', resets the counter, and generates a Type D frame at the circuit output. After this, the state machine waits for 6 clocks to fully load the local register, by asserting *SEL* to '5' and increasing the counter by '1' for every clock. Similar operations take place for all other frame types. When *RESYNC_EN* is asserted, the state machine always chooses Type E frames by asserting *SEL* to '4'. For Type E frames, original samples instead of error samples are used.

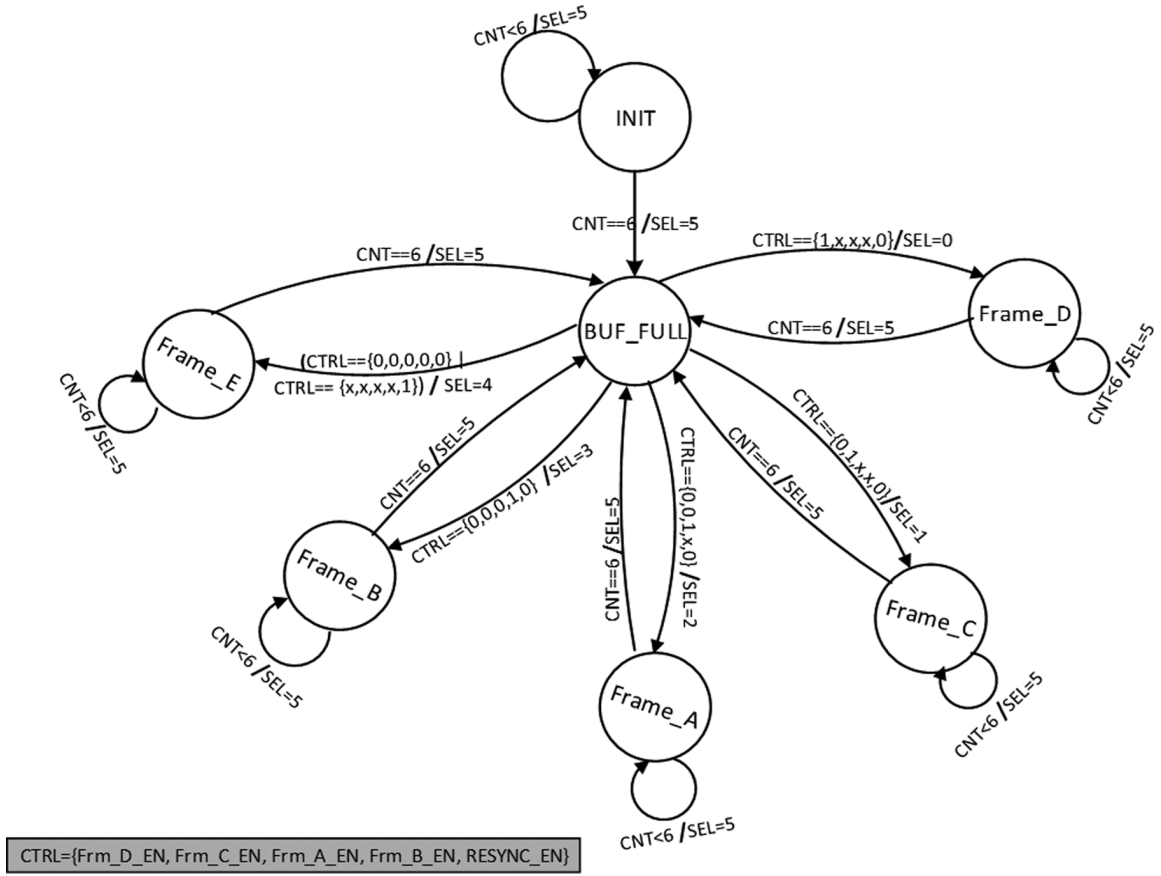

Fig. 17. Framing controller state machine.

TABLE IV
CHIP PERFORMANCE SUMMARY

| | |
|---|---|
| Supply Voltage | 2.4 V |
| Technology | 0.35 μm CMOS |
| High-Pass Frequency | 0.0075 Hz |
| Low-Pass Frequency | 35 ~ 175 Hz |
| Pass-Band Gain | 47, 54, 61, 66 dB |
| Input-Referred Noise (0.5 ~ 250 Hz) | 1.46 μVrms |
| Noise Efficiency Factor (NEF) | 3.31 |
| Common-Mode Rejection Ratio (CMRR) | 65 dB |
| Power Supply Rejection Ratio (PSRR) | 76 dB |
| Total Harmonic Distortion (THD) | -64 dBFS |
| Sampling Frequency | 256/512 Hz |
| Total Front-End Current | 12.5 μA |
| Total Back-End Current | 0.89 μA |
| ADC Effective Number of Bits (ENOB) | 9.3 |

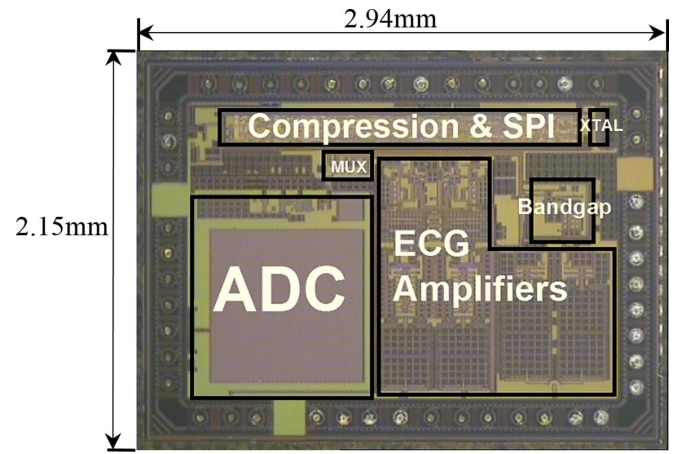

Fig. 18. Die photograph.

## VI. MEASUREMENT RESULTS AND COMPARISON

The design is implemented in a standard 0.35 μm CMOS process. The measurement is performed at 2.4 V and 3.0 V. The measured results at 2.4 V are shown in Table IV. The total input-referred noise integrated from 0.5 Hz to 250 Hz is less than 1.5 $\mu V_{rms}$. The high-pass corner is less than 0.01 Hz, and the full-scale 3 V THD is only 0.08%. The total front-end current, including all four ECG channels, ADC, DRL, bandgap, and crystal oscillator circuit, is 12.5 $\mu A$. The digital back-end consumes 0.89 $\mu A$ and has a core area of 0.2 $\times$ 2.0 mm$^2$. Fig. 18 shows the die photograph. The total chip area is 2.94 $\times$ 2.15 mm$^2$. Fig. 19 gives an example of the compressed signal and the ECG signal reconstructed from it. The signal tapped out from the ADC is overlapped with the reconstructed data and no data losses are observed in the decompressed ECG signal. Fig. 20 shows the evaluation prototype and gateway application developed to monitor ECG.

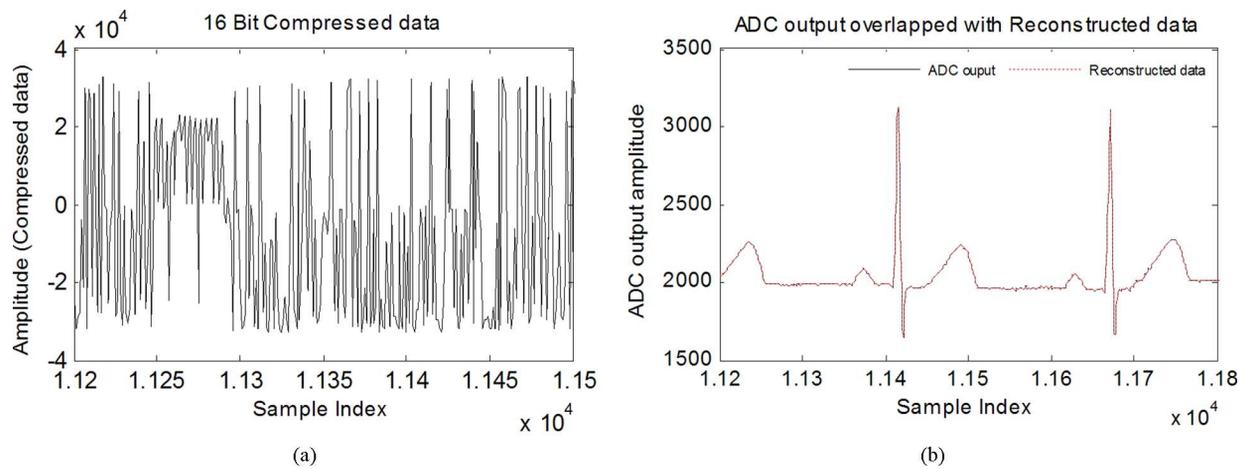

Fig. 19. (a) Transmitted compressed data. (b) ADC output ECG overlapped with reconstructed data.

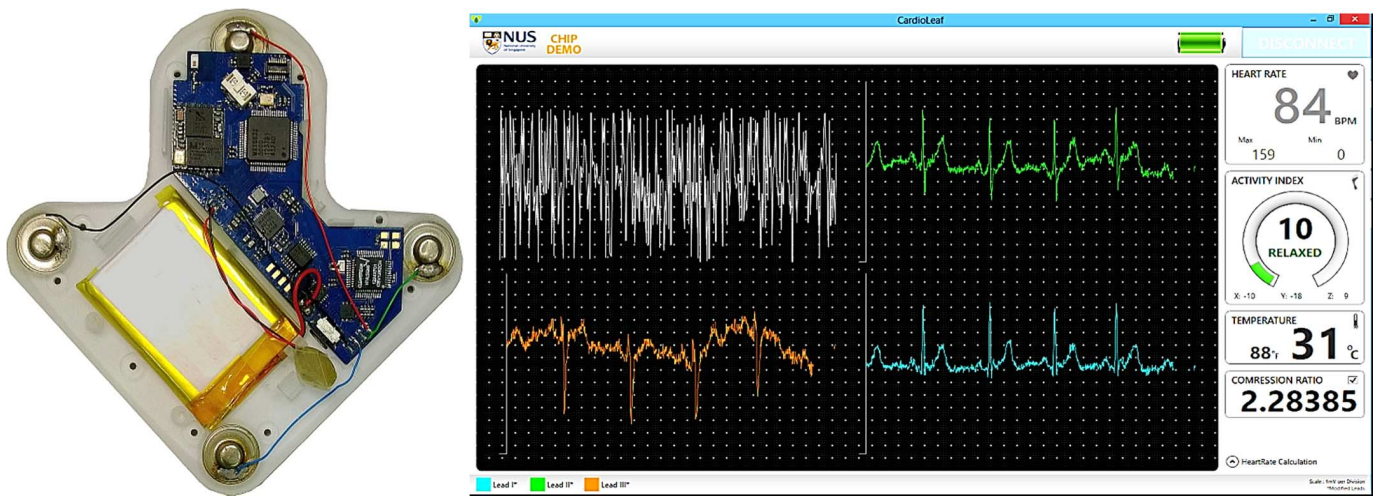

Fig. 20. Prototype device and gateway application.

Fig. 21 shows the input-referred noise spectrum of the ECG channel. The total input-referred noise is 1.47 $\mu V rms$, integrated from 0.5 Hz to 250 Hz. The thermal noise floor is at 55 nV/$\sqrt{Hz}$, with the 1/f noise corner at around 50 Hz. This design achieves a NEF of 3.31.

The digital back-end was measured under sampling frequency of 512 Hz. The compression block was operated at 32 KHz and was clock gated by the ADC's "end of conversion" signal. SPI readout was operated at 2 MHz. Since the chip includes an analog front-end, the compression performance was measured using a ST-Electromedicina ST-10 ECG signal generator. The measurement shows that the chip achieves an

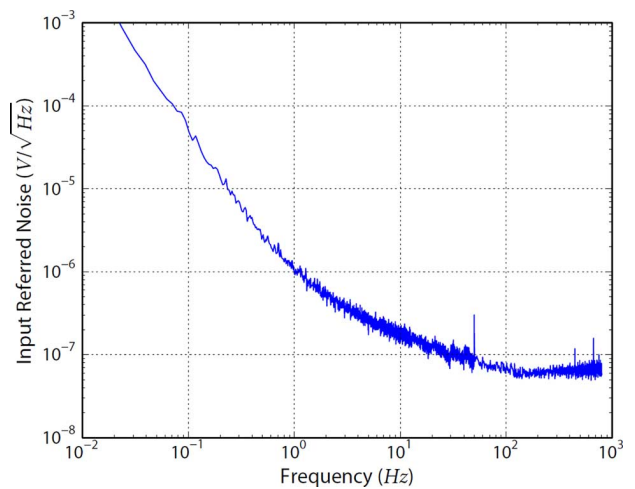

Fig. 21. Input-referred noise of the AFE.

average compression rate of 2.55 under different heart rate conditions. Comparison of the proposed compressor with other recently published designs is given in Table V. This design achieves the lowest power and complexity with a negligible reduction in compression ratio. As the compression ratio varies with sampling rate, for a fair comparison, we present two sets

TABLE V
COMPARISON OF COMPRESSOR WITH OTHER DESIGNS

|  | E.Chua [11] 2011 | S. L. Chen [17] 2013 | This design |
|---|---|---|---|
| Process | 65 nm | 0.18 μm | 0.35 μm |
| Supply Voltage (V) | 1.0 | 1.8 | 2.4 |
| Operation frequency | 24MHz | 100MHz | 32 KHz / 2MHz |
| Functions Supported | ECG/EEG/DOT | ECG | ECG |
| Compression Ratio (Simulated based on MIT Database) | 2.38 | 2.43 | 2.25 |
| Compression Ratio (Measured @ 512Hz using ECG calibrator) | - | - | 2.55 |
| Number of Channels | 3 ECG/4 EEG/1 DOT | 1 ECG | 4 ECG |
| Total Gate Count | 53.9 K | 3.57 K | 2.26 K |
| Gate Count/Channel | - | 3.57 K | 0.56 K |
| Total Power (μW) | 170* | 36.4* | 2.14 |
| Power/Channel (μW) | - | 36.4 | 0.535 |

*Simulated

of compression ratios in Table V, i.e., simulated results based on the MIT/BIH database and measured results at a sampling frequency of 512 Hz. For simulation results, all ECG signals are sampled at 360 Hz before compression and all compressors are tested with the same MIT/BIH arrhythmia database. The design in [11] gives 5.8% higher compression ratio at the cost of 23 times more gate counts due to the high memory requirements of the Golomb coding scheme. Note that the design in [11] also supports 4 EEG channels and 1 DOT operation. The selective Huffman coding scheme in [17] achieves 8% more compression at the cost of 6.4 times more of gate counts. Also noted is that only 9 bits are used in [17] to represent the prediction error, $e(n)$, for an 11 bit input data, when the data is left uncoded. Based on our understanding, this will result in data loss and therefore may not be considered fully lossless. For a fully lossless representation, $e(n)$ has to be represented by at least 13 bits. In our study, we obtained 2.15× compression using the selective Huffman coding scheme with a full bit-width representation for uncoded signals, as shown in Table III. The power consumption of the proposed design is 535 nW for 1 channel while the design in [17] consumes 36.4 μW for 1 channel. Furthermore, the output generated by [17] has variable bit lengths and has to be further packaged into fixed length for practical interfacing purposes, which involves further hardware cost [11], [14]. Overall, the proposed design gives the lowest power and gate count for compression without making any compromises on data integrity.

## VII. CONCLUSION

This paper presents a low-power ECG SoC with lossless data compression for wearable devices. The compressor achieves an average CR of 2.25× using MIT/BIH test data. The design consumes 535 nW/channel and has a core area of 0.4 mm² in 0.35 μm process. In comparison with existing methods, the proposed algorithm and hardware implementation demonstrates the lowest power consumption and is therefore suitable for wearable wireless devices.

## REFERENCES

[1] X. Zou, X. Xu, L. Yao, and Y. Lian, "A 1-V 450-nW fully integrated programmable biomedical sensor interface chip," *IEEE J. Solid-State Circuits*, vol. 44, no. 4, pp. 1067–1077, Apr. 2009.

[2] C. J. Deepu, X. Xu, X. Zou, L. Yao, and Y. Lian, "An ECG-on-chip for wearable cardiac monitoring devices," in *Proc. 5th IEEE Int. Symp. Electron. Des. Test Appl.*, 2010, pp. 225–228.

[3] F. Yazicioglu, S. Kim, T. Torfs, H. Kim, and C. Van Hoof, "A 30 μW analog signal processor ASIC for portable biopotential signal monitoring," *IEEE J. Solid-State Circuits*, vol. 46, no. 1, pp. 209–223, Jan. 2011.

[4] C. J. Deepu, X. Zhang, W.-S. Liew, D. L. T. Wong, and Y. Lian, "An ECG-SoC with 535 nW/channel lossless data compression for wearable sensors," in *Proc. 2013 IEEE Asian Solid-State Circuits Conf. (A-SSCC)*, 2013, pp. 145–148.

[5] R. R. Harrison and C. Charles, "A low-power low-noise CMOS amplifier for neural recording applications," *IEEE J. Solid-State Circuits*, vol. 38, no. 6, pp. 958–965, Jun. 2003.

[6] Y. Tsividis and C. McAndrew, *Operation and Modeling of the MOS Transistor*, 3rd ed. Oxford, U.K.: Oxford Univ. Press, 2012.

[7] M. S. J. Steyaert and W. M. C. Sansen, "A micropower low-noise monolithic instrumentation amplifier for medical purposes," *IEEE J. Solid-State Circuits*, vol. SSC-22, no. 6, pp. 1163–1168, 1987.

[8] C. Enz, F. Krummenacher, and E. Vittoz, "An analytical MOS transistor model valid in all regions of operation and dedicated to low-voltage and low-current applications," *Analog Integr. Circuits Signal Process.*, vol. 8, no. 1, pp. 83–114, 1995.

[9] J. Ramirez-Angulo, A. J. Lopez-Martin, R. G. Carvajal, and F. M. Chavero, "Very low-voltage analog signal processing based on quasi-floating gate transistors," *IEEE J. Solid-State Circuits*, vol. 39, no. 3, pp. 434–442, Mar. 2004.

[10] X. Zou, W.-S. Liew, L. Yao, and Y. Lian, "A 1 V 22 μW 32-channel implantable EEG recording IC," in *2010 IEEE Int. Solid-State Circuits Conf. (ISSCC) Dig. Tech. Papers*, 2010, pp. 126–127.

[11] E. Chua and W. Fang, "Mixed bio-signal lossless data compressor for portable brain-heart monitoring systems," *IEEE Trans. Consum. Electron.*, vol. 57, no. 1, pp. 267–273, 2011.

[12] A. Jas, J. Ghosh-Dastidar, M.-E. Ng, and N. A. Touba, "An efficient test vector compression scheme using selective Huffman coding," *IEEE Trans. Comput. Des. Integr. Circuits Syst.*, vol. 22, no. 6, pp. 797–806, Jun. 2003.

[13] S. Rigler, W. Bishop, and A. Kennings, "FPGA-based lossless data compression using Huffman and LZ77 algorithms," in *2007 Proc. Can. Conf. Electr. Comput. Eng.*, 2007, pp. 1235–1238.

[14] R. A. Becker and T. Acharya, "Variable length coding packing architecture," U.S. Patent 6653953 B2, Nov. 25, 2003.

[15] MIT-BIH Arrhythmia Database. [Online]. Available: http://physionet.org/physiobank/database/mitdb/

[16] MIT-BIH ECG Compression Test Database. [Online]. Available: http://physionet.org/physiobank/database/cdb/

[17] S.-L. Chen and J.-G. Wang, "VLSI implementation of low-power cost-efficient lossless ECG encoder design for wireless healthcare monitoring application," *Electron. Lett.*, vol. 49, no. 2, pp. 91–93, Jan. 2013.

[18] D. R. Zhang, C. J. Deepu, X. Y. Xu, and Y. Lian, "A wireless ECG plaster for real-time cardiac health monitoring in body sensor networks," in *2011 Proc. IEEE Biomedical Circuits and Systems Conf. (BioCAS)*, 2011, pp. 205–208.

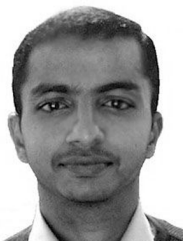

**Chacko John Deepu** (S'07–M'14) received the B.Tech. degree in electronics and communication engineering from the University of Kerala, India, in 2002, and the M.Sc. and Ph.D. degrees in electrical engineering from the National University Singapore (NUS), Singapore, in 2008 and 2014, respectively.

He is now a Research Fellow with the Bioelectronics Laboratory, NUS. His research interests include biomedical signal processing, low-power design and wearable devices.

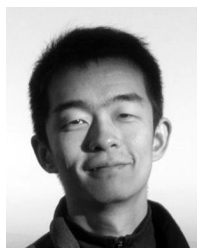

**Xiaoyang Zhang** (S'09) received the B.Sc. degree in microelectronics from Peking University, Beijing, China, in 2009. Currently he is working toward the Ph.D. degree in electrical engineering at the National University of Singapore.

His research interests include low-power biomedical sensors and wearable computing.

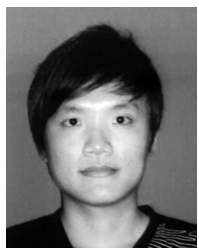

**Wen-Sin Liew** (S'08–M'13) received the B.Eng. and Ph.D. degrees in electrical engineering from the National University of Singapore, Singapore, in 2007 and 2013, respectively.

He was with the VLSI and Signal Processing Laboratory and the Bioelectronics Laboratory, National University of Singapore, from 2007 to 2013. Since 2014, he has been with Avago Technologies, Singapore, working on high-speed SerDes design. His interests include analog-to-digital converters, mixed-signal circuits, biomedical sensor interfaces, and high-speed SerDes design.

Dr. Liew was the winner of 2010 DAC/ISSCC Student Design Contest Award.

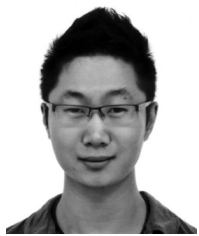

**David Liang Tai Wong** (S'11) received the B.Eng. degree with honors in computer systems engineering from Curtin University of Technology, Western Australia, in 2007, and the M.Eng. degree in integrated circuits and embedded systems from the National University of Singapore, Singapore, in 2014.

He was with Panasonic as a research and development engineer/technical lead from 2008 to 2011. He is currently a research engineer with the National University of Singapore. His research interests include low-power embedded hardware and software co-design, data communications and networks.

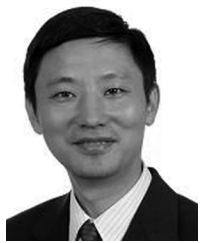

**Yong Lian** (M'90–SM'99–F'09) received the B.Sc. degree from the College of Economics and Management, Shanghai Jiao Tong University, Shanghai, China, in 1984, and the Ph.D. degree from the Department of Electrical Engineering, National University of Singapore (NUS), Singapore, in 1994.

He spent nine years in industry, and joined the NUS in 1996. He was appointed as the first Provost's Chair Professor in the Department of Electrical and Computer Engineering of NUS in 2011. He is also the Founder of ClearBridge VitalSigns Pte. Ltd., a start-up for wireless wearable biomedical devices. His research interests include biomedical circuits and systems and signal processing.

Dr. Lian has received several awards, including the 1996 IEEE CAS Society's Guillemin-Cauer Award for the best paper published in the IEEE TRANSACTIONS ON CIRCUITS AND SYSTEMS II, the 2008 Multimedia Communications Best Paper Award from the IEEE Communications Society for the paper published in the IEEE TRANSACTIONS ON MULTIMEDIA, the 2011 IES Prestigious Engineering Achievement Award, the 2012 Faculty Research Award, and the 2014 CN Yang Award in Science and Technology for New Immigrant (Singapore). As an educator, he received the University Annual Teaching Excellent Award in two consecutive academic years from 2008 to 2010 and many other teaching awards from the Faculty of Engineering. Under his guidance, his students have received many awards, including the Best Student Paper Award in ICME 2007, winner of 47th DAC/ISSCC Student Design Contest in 2010, and Best Design Award in the A-SSCC 2013 Student Design Contest.

Dr. Lian is the Vice President for Publications of the IEEE Circuits and Systems (CAS) Society, Steering Committee Member of the IEEE TRANSACTIONS ON BIOMEDICAL CIRCUITS AND SYSTEMS and the IEEE TRANSACTIONS ON MULTIMEDIA, and Past Chair of DSP Technical Committee of the IEEE CAS Society. He was the Editor-in-Chief of the IEEE TRANSACTIONS ON CIRCUITS AND SYSTEMS II: EXPRESS BRIEFS for two terms from 2010 to 2013. He also served as Associate Editor for the IEEE TRANSACTIONS ON CIRCUITS AND SYSTEMS I: REGULAR PAPERS, IEEE TRANSACTIONS ON CIRCUITS AND SYSTEMS II: EXPRESS BRIEFS, IEEE TRANSACTIONS ON BIOMEDICAL CIRCUITS AND SYSTEMS, and *Journal of Circuits, Systems Signal Processing* in the past 15 years, and was Guest Editor for eight special issues in IEEE TRANSACTIONS ON CIRCUITS AND SYSTEMS I: REGULAR PAPERS, IEEE TRANSACTIONS ON BIOMEDICAL CIRCUITS AND SYSTEMS, and *Journal of Circuits, Systems Signal Processing*. He was the Vice President for the Asia-Pacific Region of the IEEE CAS Society from 2007 to 2008, AdComm Member of the IEEE Biometrics Council from 2008 to 2009, CAS Society Representative to the BioTechnology Council from 2007 to 2009, Chair of the BioCAS Technical Committee of the IEEE CAS Society from 2007 to 2009, member of the IEEE Medal for Innovations in Healthcare Technology Committee from 2010 to 2012, and a Distinguished Lecturer of the IEEE CAS Society from 2004 to 2005. He is the Founder of the International Conference on Green Circuits and Systems, the Asia-Pacific Conference on Postgraduate Research in Microelectronics and Electronics, and the IEEE Biomedical Circuits and Systems Conference. He is a Fellow of the Academy of Engineering Singapore.